\documentstyle[11pt,mrs2003,epsfig]{article}
\begin{document} 
\title{CAUSAL FIELD EQUATIONS AND REAL EIGENVALUES 
FROM A NON-LOCAL LAGRANGIAN}

\author{Marc Soussa}
\affil{University of Florida}

\begin{abstract} 
Recently, we proposed a non-local relativistic formulation of MOND
(Modified Newtonian Dynamics) \cite{Soussa1}. The equations
of motion were not derived, rather they were inferred from the result
one would obtain by using the Schwinger-Keldysh formalism.  The
formalism simultaneously ensures causality of the field equations and
the reality of in-out operator amplitudes.  This point was avoided in
\cite{Soussa1} as its discussion was too far afield.  Here we
first demonstrate the features non-local actions generally possess: namely
acausal equations of motion and non-real in-out operator amplitudes; and
secondly how the Schwinger-Keldysh formalism works to provide the
characteristics we usually desire from effective theories.
\end{abstract} 
 
\section{Introduction} 
Previous relativistic extensions of MOND centered upon
scalar-tensor theories \cite{Milgrom} thereby introducing additional
degrees of freedom.  We instead focused on a purely metric
formulation and thus considered only gravitational degrees of freedom.
A purely metric versus a scalar-tensor approach possesses
a very different interpretation of the metric.
For the class of theories presented in \cite{Milgrom} a distinction
was made between a ``gravitational'' and ``physical'' metric 
(see \cite{Bekenstein} for a complete treatment).  The
former is responsible for gravitational dynamics whereas the latter
determines particle geodesics.  Of course, a purely
metric approach makes no distinction between ``gravitational'' and
``physical'' metric.  They are identical and thus the strong
equivalence principle is in play.  

MOND was proposed by Milgrom in 1983 to offer an alternative to the
dark matter description of the rotation curve
phenomenon -- by altering gravity at low
acceleration scales, one could reproduce the asymptotically constant
velocities of satellites outside the central galactic bulge
\cite{Milgrom}.  Although a non-relativistic action
principle was constructed which possessed the usual symmetries we seek in
mechanical systems, no obvious relativistic extension presented itself.
The main motivation for undertaking this challenge of formulating a 
relativistic version of MOND was to answer the gravitational lensing
problem.  Alone, General Relativity is unable to account for the
amount of observed galactic lensing without invoking dark matter
\cite{Mortlock}.  If MOND is to be considered an \textit{alternative}
to dark matter, it must have an impact on lensing.  We began the
project with every intention and expectation of arriving at a 
phenomenologically viable theory.
Toward the end, however, we came to the conclusion
that \textit{no} purely metric formulation of MOND would be able to
produce sufficient lensing (see \cite{Soussa2} for the assumptions
and analysis which led to this no-go statement).  Regardless of how 
one views this result, there is still the question: can a relativistic,
phenomenologically viable theory of MOND be formulated?  The
scalar-tensor approaches have proven awkward when lensing constraints
are taken into account \cite{Bekenstein2}.  Although the class of
models we considered
demonstrated this phenomenological ``disaster'', the formalism
involved for deriving the field equations is quite instructive and
will be relied upon in calculating the correction to Newton's law of
gravity in a locally de Sitter background due to quantum effects
\cite{Soussa3}.  

One interested in simplicity, viz. by pure degrees of freedom
would certainly want to consider a purely metric extension of MOND, if
for no other reason than theoretical completeness.  
However, as is usually the case in physics, one often
gains simplicity in one facet of a theory only to lose it in another.
In our case, the ``sacrifice'' we had to make was locality.  The
reason rests upon requirements placed on the Newtonian potential in
the MOND limit.  For MOND to faithfully reproduce the rotation curve
data (the very \textit{raison d'\^etre} for MOND) it must be that in
the non-relativistic regime the Newtonian potential satisfies,
\begin{equation}
\vec{\nabla}\cdot\left[\mu\left(\frac{\Vert\vec{\nabla}\phi_{\scriptscriptstyle
N}\Vert}{a_0}
\right)\vec{\nabla}\phi_{\scriptscriptstyle N}\right]=4\pi G\rho_m. 
\label{nrmond}
\end{equation}
Here $\mu(x)$ is to be considered an interpolating function
constructed to possess the proper MOND limiting behavior, namely 
$\mu(x) \to x \ \ \forall \ \ x \ll 1$.  This corresponds to accelerations 
$a \sim a_0 \sim 10^{-10}\,\rm{m/s^2}$.
Now consider the weak-field expansion of General Relativity
about a Minkowski background,
\begin{equation}
S=\frac{1}{16\pi G}\int d^4x\sqrt{-g}R \longrightarrow \frac{1}{16\pi
  G} \int d^4x\left\{h_{\mu\nu}^{\ \ ,\mu\nu}-h_{,\mu}^{\ \ \mu}+
  {\mathcal O}(h^2) \right\}.
\end{equation}
It does not take long before one realizes that there are no
\textit{local}, invariant curvature operators one could add to the
Einstein-Hilbert action of General Relativity to recover the
appropriate MOND behavior of (\ref{nrmond}).  Since the
non-relativistic MOND force law involves
$\Vert\vec{\nabla}\phi_{\scriptscriptstyle N}\Vert^2$, the weak-field expansion
would have to start at cubic order in the action.
However, any {\em local} curvature invariants added would start 
at least at quadratic order in the weak-field expansion
(e.g. $R^2$, $R_{\mu\nu}R^{\mu\nu}$, etc.), and thus there is no 
hope in ridding ourselves of even the linear piece. 

Abandoning locality is certainly not an uncommon phenomena anymore in
theoretical physics.  Indeed, effective theories have become more and
more commonplace in regimes where ignorance of fundamental
principles dominates.  Gravity's effective action, of course, falls in
this category; and although we are unable to even say what its full
effective action is, nothing prevents us from guessing its form.
Ockam's razor our guide, we chose the simplest class of guesses which
would be capable of satisfying the non-relativistic constraint of
equation (\ref{nrmond}),
\begin{equation}
{\mathcal L}=\frac{1}{16\pi G}\left[R+a_0^2{\mathcal F}\left(a_0^{-2}
g^{\mu\nu}\varphi_{,\mu}\varphi_{,\nu}\right)\right], \label{relmond}
\end{equation}  
where the \textit{small potential} is defined to be $\varphi[g]\equiv 
\frac{1}{\Box}R$ ($\Box$ is the covariant d'Alembertian).  The
function ${\mathcal F}(x^2)$ is chosen as was $\mu(x)$ from
(\ref{nrmond}) to have the correct limiting behavior in the different
acceleration regimes.

In general, one cannot expect (\ref{relmond}) to have causal field
equations.  It is the penalty
for varying a temporally non-local action.  A more serious concern,
however, arises when one wishes to quantize such a theory.  Non-local 
actions are indicative of effective actions, and this is the
interpretation we will adopt for this proceeding, as it was when we
considered the relativistic extension of MOND in \cite{Soussa1} and
\cite{Soussa2}.  Effective actions are related to in-out vacuum
amplitudes and therefore even if the operator we are considering is
Hermitian, we are in no way guaranteed that its matrix elements are
real.  We will show using a simple scalar field example how the 
Schwinger-Keldysh effective action restores not only 
causality to the field equations, but ensures that in the context 
of quantum mechanics we deal with real amplitudes. 

\section{A simple scalar field example} 
Consider a real, massive scalar field in four dimensions with the
action,
\begin{equation}
S_{\rm m}[\phi]=\int d^4x \, {\mathcal L}_{\rm m}(\phi,\partial_\mu\phi).
\end{equation}
We have all learned that the corresponding in-out effective action is,
\begin{equation}
\Gamma[\phi]=\frac12\int d^4y \,\phi(y)[\Box - m^2]\phi(y)-\frac12\int
d^4y\int d^4z \,\phi(y) \Pi^2(y;z)\phi(z)+ {\mathcal O}(\hbar^2),
\label{effaction}
\end{equation}
where $\Pi^2$ is the scalar self-energy operator responsible for
quantum corrections to the scalar's mass.
Functionally varying this action with respect to $\phi$ (and ignoring
surface terms) gives the
operator equation of motion to order $\hbar$,
\begin{equation}
[\Box-m^2]\phi(x)-\int d^4y \, \Pi^2(x;y)\phi(y)=0.\label{eom}
\end{equation}
Observe the non-local term of equation (\ref{eom}).  Since there is
nothing preventing $\Pi^2$ from being non-zero for spacelike
separations of its constituent spacetime points, this term is
manifestly acausal.  An analogous situation arises, of course, in
electrodynamics: the force a charged particle experiences from a static charge
distribution is instantaneous, and therefore acausal.  The problem is
resolved by using the retarded Green's function to the Lorentz
invariant wave equation.  

In addition to acausality, there is a more crucial concern: the
operator spectra.  Typically, one varies an effective action intending
to work with operator expectation values, and in particular vacuum 
expectation values if working in the Heisenberg picture.  The
effective action formalism, however, requires the matrix elements of 
an operator to be in-out amplitudes, where the in and out states exist
 at asymptotically early and late times, respectively, when any 
interactions have been turned off.  The in-out effective action is related to 
the in-out vacuum amplitude via the generating functional,
\begin{eqnarray}
W[J] &=& -i\ln \langle\Omega_{\rm{out}}|\Omega_{\rm{in}}\rangle_J,\\
\Gamma[\bar\phi] &\equiv& W[J]-J\bar\phi,
\end{eqnarray}
where, $\bar\phi \equiv \frac{\delta W[J]}{\delta J}$ and $J$ is a
source current.  Usually, the in-out amplitude is expressed in terms
of a path integral (up to multiplicative constants and measure factors
which for the purposes of this proceeding are irrelevant),
\begin{equation}
\langle\Omega_{\rm{out}}|\Omega_{\rm{in}}\rangle\propto\int{\mathcal
  D}\phi\,\exp\left\{i\int d^4x\sqrt{-g}{\mathcal
  L}_{\rm m}[\phi] \right\} \label{pathint}, 
\end{equation}
where the in and out vacua are defined on spacelike
hypersurfaces $\Sigma_i$ and $\Sigma_f$, respectively.
When the in state is identical to the out
state the matrix elements really are just expectation values.  
However, in general
there is nothing to prevent the in and out states to differ (even if
there are no source currents).  The out
vacuum is not necessarily free of particles and therefore even if we
demand the in vacuum to be such, it does not follow that at
asymptotically late times we recover our initial condition.
Consequently, we no longer expect real eigenvalues for our
operator spectrum.  
\section{The Schwinger-Keldysh Formalism}
What we would like is to work with an in-in vacuum amplitude -- i.e. a
vacuum expectation value.  Then, so long as our operators are
Hermitian, we need not worry about complex eigenvalues.  Further, we
wish to have some mechanism in place for which acausal pieces vanish
in the expectation value.  The Schwinger-Keldysh formalism is
expressly constructed to satisfy these requests.  Here we give a brief
overview.  Those interested in a more thorough 
treatment, however, are referred to \cite{Jordan}.  

It works by introducing two fields, distinguished by a plus or minus
label.  One then evolves forward from $\Sigma_i$ for which only the 
plus field $\phi_+$ is non-zero to an arbitrary 
but spacelike surface $\Sigma$, and antithetically backward from 
$\Sigma$ to $\Sigma_i$ for which only $\phi_-$ is non-zero.  
The two fields are required to satisfy the boundary
condition $\phi_+\bigl\vert_{\Sigma}=\phi_-\bigl\vert_{\Sigma}$.
This has the desired effect of transforming an in-out amplitude into an
expectation value.  The Schwinger-Keldysh generating functional is defined by
inserting a complete set of states in the presence of now two source
currents $J_\pm$,
\begin{eqnarray}
e^{iW[J_+,J_-]}&\equiv& \sum_\alpha\langle\Omega_{\rm in}|\Omega_{\rm out}^ 
\alpha\rangle_{J_-}\langle\Omega_{\rm out}^\alpha|\Omega_{\rm in}\rangle_{J_+} 
\propto\int{\mathcal D}[\phi_+]
{\mathcal D}[\phi_-]\,e^{i(S_{\rm m}[\phi_+]-S_{\rm m}[\phi_-]
+J_+\phi_+ - J_-\phi_-)}.
\end{eqnarray}
In general, one constructs time-ordered operator expectation values
by taking variations of the generating functional and setting
$J_+=J_-=J$ (it is to be understood that times along the backward
evolution are later than those along the forward evolution),
\begin{eqnarray}
\left(\frac{\delta}{i\delta J_+(x_1)}\right)
\dots \left(\frac{\delta}{i\delta J_+(x_n)}\right)
\left(\frac{\delta}{-i\delta J_-(y_1)}\right)
\dots \left(\frac{\delta}{-i\delta J_-(y_m)}\right)W[J_+,J_-] 
\Biggr\vert_{J_+=J_-=0}  \nonumber \\
  = \langle\Omega_{\rm in}|T^\dagger\{\phi(y_1)\dots\phi(y_m)\} 
T\{\phi(x_1)\dots\phi(x_n)\}|\Omega_{\rm in}
\rangle,  
\end{eqnarray}
where $T$ and $T^\dagger$ are the time and anti-time ordering symbols,
respectively.
The propagator, for instance, can be expressed as a matrix with elements,
\begin{eqnarray}
i\Delta_{++}(x,y) &=& i\langle\Omega_{\rm
  in}|T\{\phi(x)\phi(y)\}|\Omega_{\rm in} 
\rangle \label{spp}\\
i\Delta_{+-}(x,y) &=& i\langle\Omega_{\rm
  in}|\phi(y)\phi(x)|\Omega_{\rm in} \rangle, 
\label{spm}\\
i\Delta_{-+}(x,y) &=& i\langle\Omega_{\rm in}|\phi(x)\phi(y)|\Omega_{\rm
  in}\rangle \label{smp}\\
i\Delta_{--}(x,y) &=& i\langle\Omega_{\rm in}|T^\dagger\{\phi(x)\phi(y)\}
|\Omega_{\rm in}\rangle.\label{smm}
\end{eqnarray}
Working to order $\hbar$, the Schwinger-Keldysh effective action 
(or the in-in effective action) is,
\begin{eqnarray}
\Gamma[\bar\phi_+,\bar\phi_-]&\equiv& W[J_+,J_-]-J_+\bar\phi_+ - J_-\bar\phi_-,
\nonumber \\
&=& S_{\rm m}[\bar\phi_+]-S_{\rm m}[\bar\phi_-]-\frac12\int d^4y
\int d^4z \bar\phi_+(y)\Pi^2_{++}(y;z)\bar\phi_+(z)  \nonumber \\
& &- \frac12\int d^4y\int d^4z \bar\phi_+(y)\Pi^2_{+-}(y;z)\bar\phi_-(z)
- \frac12\int d^4y\int d^4z \bar\phi_-(y)\Pi^2_{-+}(y;z)\bar\phi_+(z) 
\nonumber \\
& &+ \frac12\int d^4y\int d^4z \bar\phi_-(y)\Pi^2_{--}(y;z)\bar\phi_-(z),
\label{skaction}
\end{eqnarray}
where now $\bar\phi_\pm \equiv \pm \frac{\delta W[J_+,J_-]}{\delta
  J_\pm}$.  The self-energy operators satisfy,
\begin{eqnarray}
\Pi^2_{++}(x;y) &=& \Pi^2_{++}(y;x) \qquad,\qquad \Pi^2_{--}(x;y) = 
\Pi^2_{--}(y;x),\\
\Pi^2_{+-}(x;y) &=& \Pi^2_{-+}(y;x).
\end{eqnarray}
In addition to the $\pm$ labels, the operators are related by simple
rules which alter the $i\epsilon$ terms in propagators and change the
sign of some vertices.
One obtains the field equations by varying (\ref{skaction}),
\begin{equation}
\frac{\delta\Gamma[\bar\phi_+,\bar\phi_-]}{\delta\bar\phi_+} 
\Biggr\vert_{\bar\phi_+ =\bar\phi_- =\phi}
= [\Box-m^2]\phi(x) -\int d^4y
\left[\Pi^2_{++}(x;y)+\Pi^2_{+-}(x;y)\right] \phi(y). \label{effequation}
\end{equation}
It is not readily obvious that equation (\ref{effequation})
is causal and pure real.  In (\ref{effequation}), these properties are
realized via the relations,
\begin{eqnarray}
\Pi^2_{+-}(x;y) &=& -\Pi^2_{++}(x;y) \quad \forall \quad x,y \ \textnormal{
spacelike separated} \label{causal},\\
\Pi^2_{+-}(x;y) &=& \Pi^{2*}_{++}(x;y) \quad \ \ \, \forall \quad x,y \ \
\textnormal{timelike separated}. \label{real}
\end{eqnarray}
One can easily verify that (\ref{causal})-(\ref{real}) are satisfied
by the propagator equations (\ref{spp})-(\ref{smm}).  For an example
of this procedure and a more detailed treatment of the technical
aspects which arise during calculations, the reader is referred to
\cite{Prokopec}.
    
\section{Conclusions}
Deriving MOND from purely gravitational principles forces one to consider
non-local actions.  The result is that causality of the field
equations and reality of the operator spectra are no longer ensured.  
Using a real scalar field example, we have shown how the Schwinger-Keldysh
formalism ovecomes these two obstacles.  The result is
causal and real field equations.  

In our example, the Schwinger-Keldysh self-energy operators satisfied
certain relationships with
each other dependant upon the separation of their constituent
spacetime points.  If spacelike, they were exactly minus each other
thus ensuring causality.  If timelike, they were complex conjugates
and therefore the sum real.  These properties ultimately derive from
the vertex signs and the regularization adopted
but not shown in (\ref{spp})-(\ref{smm}).  

The analysis here is easily adapted
to higher spin theories, and it is the intent to use this formalism to
calculate the quantum corrections to $G_N$ in a locally de Sitter
background \cite{Soussa3}.  Although we did not concern ourselves with
quantizing the relativistic MOND theory, it is important to not only have an
understanding of what the quantum implications of adding operators to
classical actions have but also to have a formalism that generates for us
physically meaningful quantities.  

\acknowledgements{ 

I thank Richard Woodard for his help and encouragement.  
}

\vfill 
\end{document}